\documentclass[pre,aps,superscriptaddress]{revtex4}
\usepackage{amsmath}
\usepackage{graphicx}

\def\mc{\mathcal}
\def\be{\begin{equation}}
\def\ee{\end{equation}}
\def\bc{\begin{center}}
\def\ec{\end{center}}
\def\bea{\begin{eqnarray}}
\def\eea{\end{eqnarray}}
\def\bean{\begin{eqnarray*}}
\def\eean{\end{eqnarray*}}

\def\nn{\nonumber}

\def\log{{\rm \; log}}

\def\a{\alpha}
\def\b{\beta}

\def\hu{\hat{u}}
\def\hw{\hat{w}}
\def\hp{h^\prime}
\def\up{u^\prime}
\newcommand{\mcP}{\mathcal{P}}
\newcommand{\mcQ}{\mathcal{Q}}

\begin{document}

\title{Bicoloring random hypergraphs}

\author{Tommaso Castellani}
\affiliation{Dipartimento di Fisica, SMC and INFM, Universit\`{a} di
Roma ``La Sapienza'', Piazzale Aldo Moro 2, I-00185 Roma, Italy}
\author{Vincenzo Napolano}
\affiliation{Dipartimento di Fisica, Universit\`a di Trieste, Via
Valerio 2, I-34127 Trieste, Italy}
\author{Federico Ricci-Tersenghi}
\affiliation{Dipartimento di Fisica, SMC and INFM, Universit\`{a} di
Roma ``La Sapienza'', Piazzale Aldo Moro 2, I-00185 Roma, Italy}
\author{Riccardo Zecchina}
\affiliation{International Center for Theoretical Physics, Strada
Costiera 11, P.O. Box 586, I-34100 Trieste, Italy}

\begin{abstract}
We study the problem of bicoloring random hypergraphs, both
numerically and analytically.  We apply the zero-temperature cavity
method to find analytical results for the phase transitions (dynamic
and static) in the 1RSB approximation. These points appear to be in
agreement with the results of the numerical algorithm. In the second
part, we implement and test the Survey Propagation algorithm for
specific bicoloring instances in the so called HARD-SAT phase.
\end{abstract}

\maketitle

\section{Introduction}

The hypergraph bicoloring is one of the classic combinatorial
optimization problems belonging to the $NP$-complete class
\cite{Miller}.  Its random version, bicoloring of random hypergraphs,
is a very interesting problem for the phase transitions it
shows. Indeed, varying the average connectivity of the random
hypergraph, the model undergoes a transition \cite{chi} from a phase
in which all links can be properly colored to a phase in which a
sizeble fraction of links are violated. Around the transition point
most difficult instances accumulate.

A graph is an ensemble of sites and links between them.  In a
hypergraph, the links connect triplets of sites. Each site (or vertex)
can be colored in two ways, say black or white, so it is natural to
identify it with an Ising spin variable that can assume the values 1
or $-1$. The link is considered to be satisfied if the three spins
that share it are not all of the same color. In the following we will
often refer to a link as a function node, as it is called for example
in the K-SAT problem \cite{MPZ}. The bicoloring problem consists in
finding an assignment to all spins such that all the links are
satisfied.  Consequently a graph will be called colorable or
uncolorable.

We can write the Hamiltonian for the problem assuming that each
unsatisfied link gives a positive energy and zero otherwise.  The
total energy is proportional to the number of unsatisfied links: a
colorable hypergraph will have a zero-energy ground state, while a non
colorable one will have a positive-energy ground state.


The Hamiltonian for bicoloring a hypergraph $\mc{G}$ reads
\be \label{ham}
\mc{H}=\sum_{\{i,j,k\}\in\mc{G}}\frac{1+\sigma_i\sigma_j +
\sigma_i\sigma_k + \sigma_j\sigma_k}{2} \;,
\ee
where $\sigma_i=\pm 1$ are Ising variables (corresponding to the 2
available colors) and the sum runs over all the hyperedges of $\mc
G$. Note that a factor $2$ has been introduces for computational
convenicence~\footnote{Local fields will turn out to be integer valued
rather than fractional.}.

Each term in the above sum is equal to 2 if and only if all the spins
in the same interaction are parallel, that is if all the vertices
connected by a hyperedge have the same color. The Hamiltonian in
Eq.(\ref{ham}) thus counts twice the number of badly colored
hyperedges. Perfect colorations correspond to zero-energy
configurations.

In the present work we focus on colorability of random hypergraphs
with $N$ vertices and $M$ hyperedges, varying the relevant parameter
$\alpha=\frac{M}{N}$. In a typical random hypergraph the connectivity
of a spin (i.e. the degree of a vertex) is a random variable
distributed according to a Poissonian of mean $3\a$.

Analogously to random K-SAT \cite{MPZ}, random K-XORSAT \cite{pspin}
and Q-coloring of random graphs \cite{BMPWZ}, the random hypergraph
bicoloring is expected to undergo two phase transitions increasing
$\a$. The first one is called ``dynamical transition'' and is located
at $\a_d$ where solutions to the problem (perfect colorations) undergo
a clustering phenomenon. At this point the complexity $\Sigma$, which
counts the number of clusters of solutions, becomes non-zero.  We
remind that if $\mc{N}(E)$ is the number of states at energy $E$ the
complexity is defined by the relation
$\mc{N}(E)=\exp{N\Sigma(\a,E/N)}$, so it is a function of $\a$ and of
the energy density. In the region where the complexity becomes
positive, on top of a great number of ground states there appear an
even larger number of metastable states: the latter may trap and slow
down linear-time coloring algorithms and local search randomized
methods \cite{blrz}.  At present all known linear-time coloring
algorithms stop converging for $\alpha$ values well below $\alpha_d$.

The second transition takes place at $\a_c$, where the ground-state
energy becomes positive: for $\a<\a_c$ most of the hypergraphs are
colorable, while for $\a>\a_c$ most of them are not. This transition
is formally equivalent to the so called SAT/UNSAT transition of K-SAT
\cite{MPZ,ksat} and K-XORSAT \cite{pspin}, and we will refer to it
with this name, although it is also known as ``COL/UNCOL'' transition
in the computer science literature.

Known results on the SAT/UNSAT transition are only upper and lower
bounds. The best upper bound for $\a_c$, found with rigorous
calculation, is 2.409 \cite{alon}. The best lower bound is 3/2
\cite{Dimitris1}.  In Ref.\cite{Dimitris2} it is analyzed the more
general problem of bicoloring random hypergraphs with $p$-spin
hyperlinks.  However for the $p=3$ case the bounds are worse than the
ones we mentioned above.  Recent rigorous results on random spin
models and random K-SAT (K even) \cite{GuerraToninelli,FranzLeone}
have shown that the 1RSB results provide rigorous upper bounds to the
phase transition point and we expect the same to be true in our case.

\section{Numerical results}

\begin{figure}
\begin{center}
\includegraphics[width=.49\textwidth]{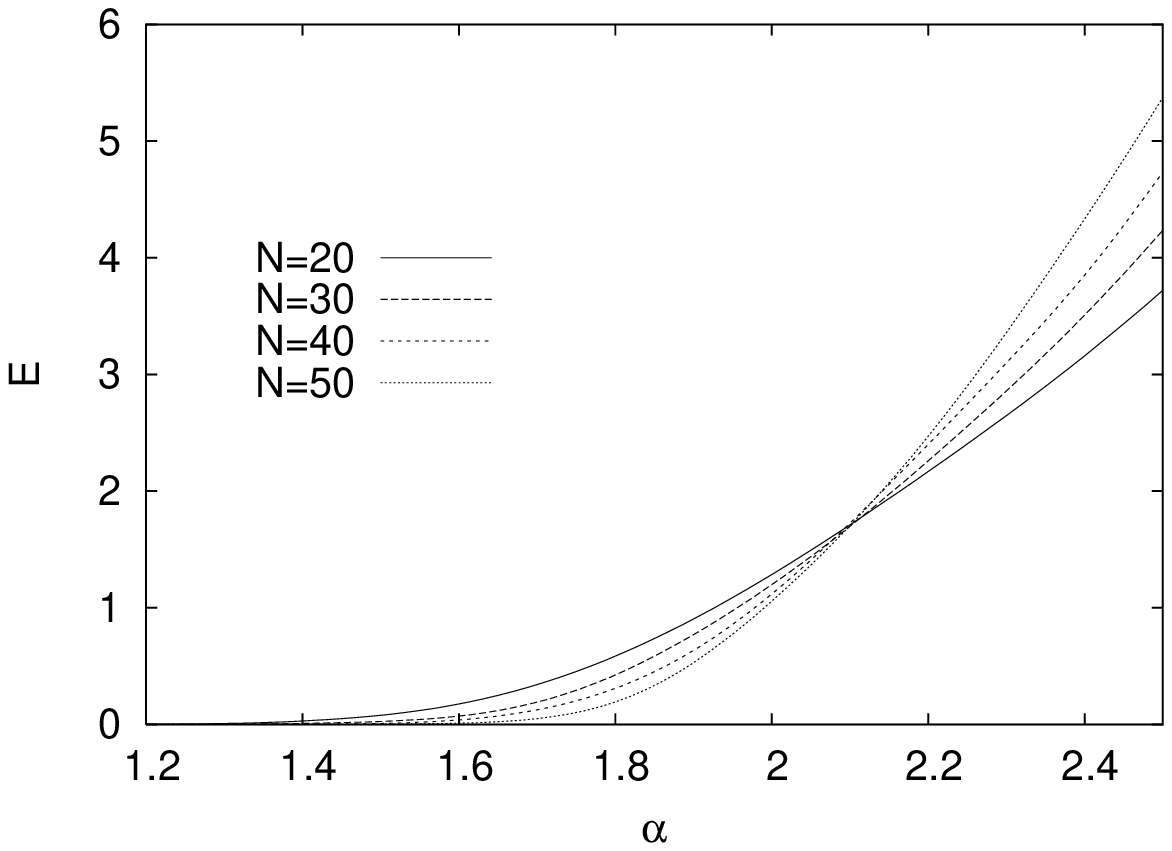}
\includegraphics[width=.49\textwidth]{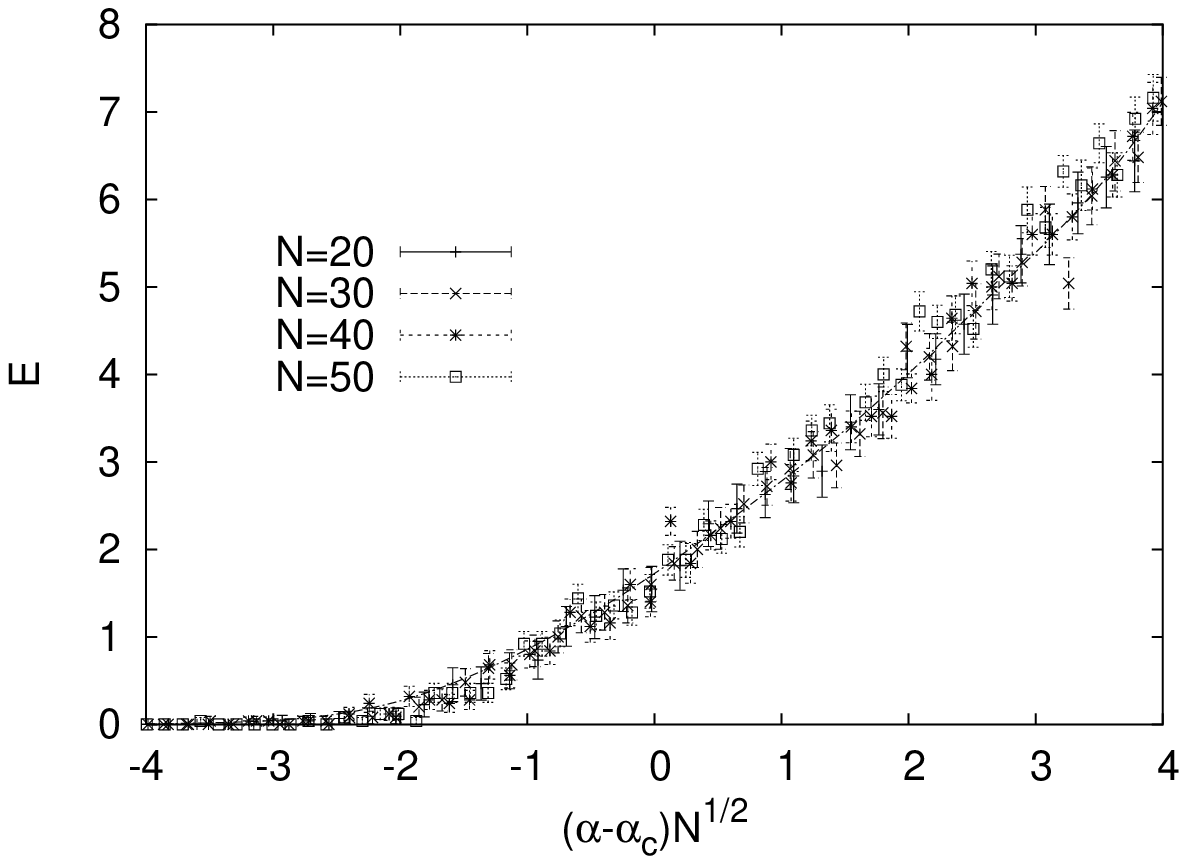}
\end{center}
\caption{Left: Average extensive energy for sizes $N=20,30,40,50$. The
crossing point roughly localizes the SAT/UNSAT transition.  Right:
Average extensive energy as a function of the rescaled variable
$(\a-\a_c)N^{1/2}$. Data are represented with standard deviations.}
\label{e}
\end{figure}

\begin{figure}
\begin{center}
\includegraphics[width=.5\textwidth]{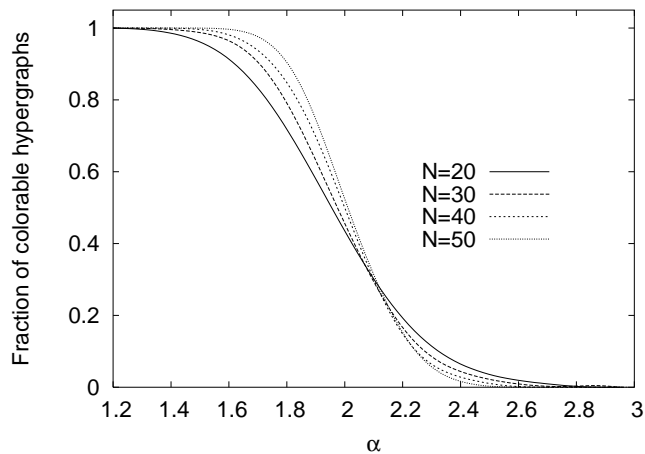}
\end{center}
\caption{Fraction of colorable hypergraphs at $N=20,30,40,50$. The
finite-size corrections are in this case larger and then the crossing
point is less clearly localized.}
\label{frac}
\end{figure}

We wrote a recursive Davis-Putnam algorithm \cite{brian} to color
random finite-size hypergraph in order to localize the point $\a_c$,
that will be calculated analytically in the next sections.  Here we
present the numerical results, whose uncertainties are very small
thanks to the average of a large number of disorder realizations.  In
Fig.~\ref{e} (left) we show that the energy curves for different $N$
cross at $\a_c$. Indeed for $\a<\a_c$ $\lim_{N\rightarrow\infty}E=0$
because all hypergraphs are colorable, while for $\a>\a_c$ $E\propto
N$ and diverges for $N\rightarrow\infty$. From Fig.~\ref{e} we
estimate $\a_c\simeq 2.1$.  All the curves can be nicely collapse when
plotted versus $(\a-\a_c)N^{1/2}$, see Fig.~\ref{e} (right).

A second estimate of $\a_c$ can be obtained from the curves of the
probability of being colorable as a function of $\a$ (see
Fig.~\ref{frac}). However here the crossing point is less clear
because of larger finite-size corrections.

\section{The cavity replica symmetric solution}
\subsection{Self-consistency equations}

We now study the bicoloring problem with the cavity method at zero
temperature \cite{bethe,cavity}.  The simplest form of the
zero-temperature cavity method is the Replica Symmetric (RS)
approximation, in which we suppose the system to have a single state.
The basic hypothesis of the cavity method is the lack of correlation
between two randomly chosen spins, because of the local tree structure
of the hypergraph.  Thanks to these vanishing correlations, the energy
of the system for fixed $\sigma_0$ can be written as a function of the
cavity fields $h_j$ and $g_j$ on the $2k$ neighbors of $\sigma_0$
\cite{cavity}
\be
E(\sigma_0) = E_0 - \sum_{j=1}^{k}\hw(g_j,h_j) - \sigma_0
\sum_{j=1}^{k} \hu(g_j,h_j) \;.
\ee
In the case of hypergraph bicoloring the function $\hu$ and $\hw$ are
given by
\begin{equation} 
\left\{
\begin{array}{ll}
\hat{u}(h_2,h_3)=\theta(-h_2)\theta(-h_3)-\theta(h_2)\theta(h_3)\\
\hat{w}(h_2,h_3)=|h_2|+|h_3|-|u(h_2,h_3)|
\end{array} \right. \label{def}
\end{equation}
where $\theta(x)=1$ if $x>0$ and $\theta(x)=0$ elsewhere.  The $\hu$
are integers and can assume the values 0, 1 or $-1$. Note that
$\hat{w}=\sum|h|-|u|$ is a general relation for models with Ising type
variables.

In the thermodynamic limit, we can assume the probability
distributions of cavity fields $h$ and cavity biases $u$ to have well
defined limits, and write for them self-consistency equations
\begin{equation} \left\{
\begin{array}{ll}
Q(u)=\displaystyle \int dP(h_1)\,dP(h_2)\,\delta\Big(u -
\hat{u}(h_1,h_2)\Big)\;,\\
P(h)=\displaystyle \sum_{k=0}^{\infty} f_{3\alpha}(k) \int dQ(u_1)
\ldots dQ(u_k)\,\delta\Big(h-\sum_{i=1}^{k}u_i \Big) \;,
\end{array} \right.
\label{eqRS}
\end{equation}
with $$f_{3\alpha}(k)=\frac{(3\alpha)^k}{k!}e^{-3\alpha}$$ As
expected, these equations coincide with those obtained from a replica
calculation in Ref.~\cite{MOUR_SAAD}.

Exploiting system symmetries one can always write
\begin{equation}
Q(u) = c_0\,\delta(u)+\frac{1-c_0}{2}\Big[\delta(u+1)+\delta(u-1)\Big]\;.
\end{equation}
Analogously the distribution of cavity fields can be written as $P(h)
= \sum_{i=-\infty}^\infty p_i\,\delta(h-i)$, where the coefficients
$p_i$ are symmetric, i.e.\ $p_i=p_{-i}$.  The self consistency
equations can be then written in terms of $p_0$ and $c_0$ as
\begin{equation} \label{scrs} \left\{
\begin{array}{ll}
p_0=e^{-3\alpha(1-c_0)}I_0(3\alpha(1-c_0))\\
c_0=1-\frac{(1-p_0)^2}{2}
\end{array} \right.
\end{equation}
where $I_0(x)$ is the zero-order modified Bessel function.  $c_0$ is
the order parameter of the system and it satisfies the
self-consistency equation
\begin{equation} \label{rseq}
1-\sqrt{2(1-c_0)}=e^{-3\alpha(1-c_0)}I_0\Big(3\alpha(1-c_0)\Big) \;.
\end{equation}
For any $\a$ value a ``paramagnetic'' solution $c_0=1$ exists, for
which all the cavity fields are zero.  For $\a>\alpha_{RS}=2.3336$,
there also exists a non-trivial ``glassy'' solution with $c_0<1$.

\subsection{Energy density}

We now compute the RS energy density, following the notation already
used in \cite{cavity}. We must compute $E(\a)=\Delta E_1-2\alpha\Delta
E_3$ where
\begin{eqnarray}
\Delta E_3 &=& \int dP(h_1)\,dP(h_2)\,dP(h_3)\cdot \nonumber \\
&&\cdot \left[ \min_{\sigma_1,\sigma_2,\sigma_3}\left(
\frac{1+\sigma_1\sigma_2+\sigma_2\sigma_3+\sigma_1\sigma_3}{2}
-h_1\sigma_1-h_2\sigma_2-h_3\sigma_3\right)+|h_1|+|h_2|+|h_3|)\right]
\nonumber\\
&=& 2 \int dP(h_1)\,dP(h_2)\,dP(h_3)\,\theta(h_1h_2)\,\theta(h_2h_3)
=\frac{1}{2}(1-p_0)^3=\sqrt{2}(1-c_0)^{\frac{3}{2}} \;,\\
\Delta E_1 &=& \sum_{k=0}^{\infty}f_{3\a}(k) \int dQ(u_1)\ldots
dQ(u_k) \left(\sum_{i=1}^{k}|u_i|-|\sum_{i=1}^{k}u_i| \right) \nn \\
&=& 3\a(1-c_0)-2e^{-3\a(1-c_0)}\sum_{r=1}^{\infty} r
I_r\Big(3\a(1-c_0)\Big)\;.
\end{eqnarray}

If we introduce the parameter $\lambda=3\alpha(1-c_0)$ which satisfies
the equivalent of Eq.(\ref{rseq}) the total RS energy density can be
written as follows
\be \label{ers}
E = \lambda-2e^{-\lambda}\sum_{r}rI_r(\lambda) -
\frac{2}{3}\lambda\Big(1-e^{-\lambda}I_0(\lambda)\Big) \;.
\ee
The expression (\ref{ers}) seems to be the same for the different
models with Ising variables (like $p$-spin \cite{RIWEZE}, K-SAT
\cite{MZ97}, etc.), the difference being only in the self-consistency
equation for $\lambda$, where $\alpha$ is multiplied by a different
constant.  For example the $\alpha_{RS}$ value for the present
bicoloring model is twice the value it takes in the 3-spin model
\cite{RIWEZE}.

\subsection{RS phase diagram}

If we plot the energy (\ref{ers}) versus $\alpha$ we see that the
energy of the non trivial solution is negative for $\alpha<2.5906$.
In the region $2.3336<\a<2.5906$ the RS solution is therefore
non-physical, because the energy density of this problem must be
positive by definition.  In the RS approximation we have found a
paramagnetic phase for $\alpha<2.3336$ and a glassy phase for
$\a>2.5906$.  This prediction is not correct, both quantitatively and
qualitatively. The values of $\a$ where the transitions appear are not
in agreement with numerical simulations, and there is a non-physical
region.

\subsection{Instability of evanescent field in the paramagnetic
region}

Before going to the 1RSB approximation, let us concentrate in this
section on the RS paramagnetic region $\alpha<2.3336$, in order to
analyze the distribution of the so-called {\em evanescent fields}
\cite{Birolietal}.  In the paramagnetic phase at zero temperature all
the cavity fields $h_i$ are null, but considering the first order
correction in temperature one can write $h_i = T \hp_i$ (whence the
adjective evanescent).

In terms of expectation values of spin variables, an evanescent field
is the only one that can give a finite magnetization in the zero
temperature limit: $m = \tanh(\b h) \to \tanh(\hp)$. On the contrary,
in the `strictly'-zero-temperature formalism that we use to study
ground state energy, variables are either frozen, $|m|=1$, or
paramagnetic, $m=0$, and we disregard any detailed information
concerning the fluctuations of the local magnetizations of the
unfrozen variables.  The global probability distribution of the local
magnetizations could in principle be non trivial, with some variable
polarized (yet never frozen) in some preferential direction.

There are two equivalent ways of obtaining such information on the
distribution of magnetizations: The first consists in writing the
iterative cavity equations for such magnetizations and next taking the
average over the underlying random hyper-graph. The second simply
consists in computing the RS cavity equations at finite temperature
assuming appropriate scaling of the cavity fields. Taking $h_i = T
\hp_i$ with $\hp_i$ finite leads, in the $\beta \to \infty$ limit, to
a distribution of evanescent fields which may describe non trivial
expectations for the spins.

Following the same steps which brought us to the RS self-consistency
equations (\ref{eqRS}), we can write analogously the self-consistency
equations for the distributions of $\hp_i = \b h_i$ and $\up_i = \b
u_i$ in the $\b \to \infty$ limit.  These equations look identical
those in Eq.(\ref{eqRS}), the only difference being the definition of
the function $\hu(h_1,h_2)$, which now reads
\begin{equation}
\hat{u}'(\hp_1,\hp_2) = \frac{\tanh(\hp_1)+\tanh(\hp_2)}
{\tanh(\hp_1)\tanh(\hp_2)-3} \quad .
\end{equation}

For very low $\alpha$ the only solution to the self-consistency
equations is $P(h')=\delta(h')$.  At variance with respect to other
problems like for instance 3-SAT \cite{MZ97} in which the low $\alpha$
phase is highly non trivial, the bicoloring problem is simple.  As it
happens in the Q-coloring \cite{BMPWZ} and in the 3-spin problems
\cite{RIWEZE,pspin}, the very low $\alpha$ phase is a genuine
paramagnet, with local fields concentrated around zero even at the
first order in temperature.

However the solution $P(\hp)=\delta(\hp)$ and $Q(\up)=\delta(\up)$ may
become unstable at a certain value of $\a$, that we call $\a_s$.  In
order to study the stability of this solution (in which local fields
are uncorrelated independently of the local strucutre of the
underlying hypergraph) it is enough to give an infinitesimal width to
$P(\hp)$ and check whether it increases or decreases under the
iteration of Eq.(\ref{eqRS}).  For very small values of $\hp_i$ one
can linearize the function $\hat{u}'(\hp_1,\hp_2) \simeq
-(\hp_1+\hp_2)/3$ and obtain very simple relations among the variances
of $P(\hp)$ and $Q(\up)$ at two consecutive iterations ($n$ and $n+1$)
\begin{eqnarray}
\langle (\up)^2 \rangle_{n+1} &=& \frac29 \langle (\hp)^2 \rangle_n \;,\\
\langle (\hp)^2 \rangle_{n+1} &=& 3\a \langle (\up)^2 \rangle_n \;.
\end{eqnarray}

For $\a<\a_s=3/2$ the variances do not increase under iteration of the
RS equations and the system is in a truly paramagnetic phase with all
the magnetization identically zero. 

For $\a>\a_s$, the presence of a broad distribution of first-order
corrections $\hp$ suggests the presence of a full RSB spin-glass phase
at finite temperature, produced by a ``replicon'' instability at
$\a_s$.  The finite-temperature phase transition at $\a_s$ corresponds
at $T=0$ to the onset of a non trivial organization of ground states,
with non trivial magnetizations (unfrozen RSB scenario). We 
incidentally note that the 
value of $\a_s$ coincides with the best lower bound available for $\a_c$.

However, as soon as the dynamical transition is reached at $\alpha_d
\simeq 1.915$ (see next section), the system looses memory of the
unfrozen RSB phase.  The non-evanescent fields, $h=\mathcal{O}(1)$,
are the only ones relevant in determining the ground state energy.  At
the level of non vanishing fields, at $\alpha_d$ we have a transition
from RS to 1RSB.  At this point, the analytically disconnected
solution with vanishing fields disappears.  The presence of full RSB
is somehow accidental and we expect for higher number of colors to
disappear completely (as it happens in graph coloring \cite{BMPWZ}).

\section{The cavity 1RSB solution}

\subsection{Self-consistency equations: the distribution $\rho(\eta)$}

In the previous section we have seen that the RS approximations
produces a wrong solution. Here we study the system with a better
approximation, the so-called ``one step Replica Symmetry Breaking''.

In this approximation the scenario is a bit more complex: at $\a_d$
($<\a_c$) there is a clustering phenomenon so that the computation
made in the RS case is only valid within each state (cluster). It must
be also considered the crossing between the energy of two states, for
which we use the ``reweighting parameter'' $\mu$ as in \cite{cavity}.

The 1RSB order parameter is a distribution of distributions, whose
self-consistency equations are the following
\begin{eqnarray} \label{selfc}
\mcQ[Q] &=& \int D\mcP[P_1]\, D\mcP[P_2]\; \delta^{(F)}\left[ Q(u) -
\int dP_1(h_1)\, dP_2(h_2)\ \delta(u-\hat{u}(h_1,h_2)) \right] \\
\mcP[P] &=& \sum_{k=0}^{\infty} f_{3\a}(k) \int \prod_{i=1}^k
D\mcQ[Q_i]\; \delta^{(F)}\left[ P(h) - \frac{1}{A_k} \int
\prod_{i=1}^k dQ_i(u_i) e^{-\mu(\sum |u|-|\sum u|)}
\delta(h-\sum_{i=1}^{k}u_i) \right]  \label{selfc2}
\end{eqnarray}
with $\delta^{(F)}$ being a functional delta, and $A_k$ normalization
coefficients.

Thanks to the system symmetries the most general form for $Q(u)$ is
given by
\begin{equation} \label{Q}
Q(u)=\eta\, \delta(u)+\frac{1-\eta}{2}\Big[\delta(u+1)+\delta(u-1)\Big] ,
\end{equation}
that is symmetric under $u \leftrightarrow -u$ and with
$u\in\{-1,0,1\}$. The heterogeneity of the random hypergraphs is now
reflected in the very different values $\eta$ may take: e.g. isolated
plaquettes certainly have $\eta=1$. Let us call $\rho(\eta)$ the
probability distribution function of $\eta$.  The problem will be now
studied in terms of $\rho(\eta)$, which completely determines the
order parameter $\mcQ[Q]$.

\subsection{$\mu\rightarrow\infty$ limit}

\begin{figure}
\begin{center}
\includegraphics[width=.49\textwidth]{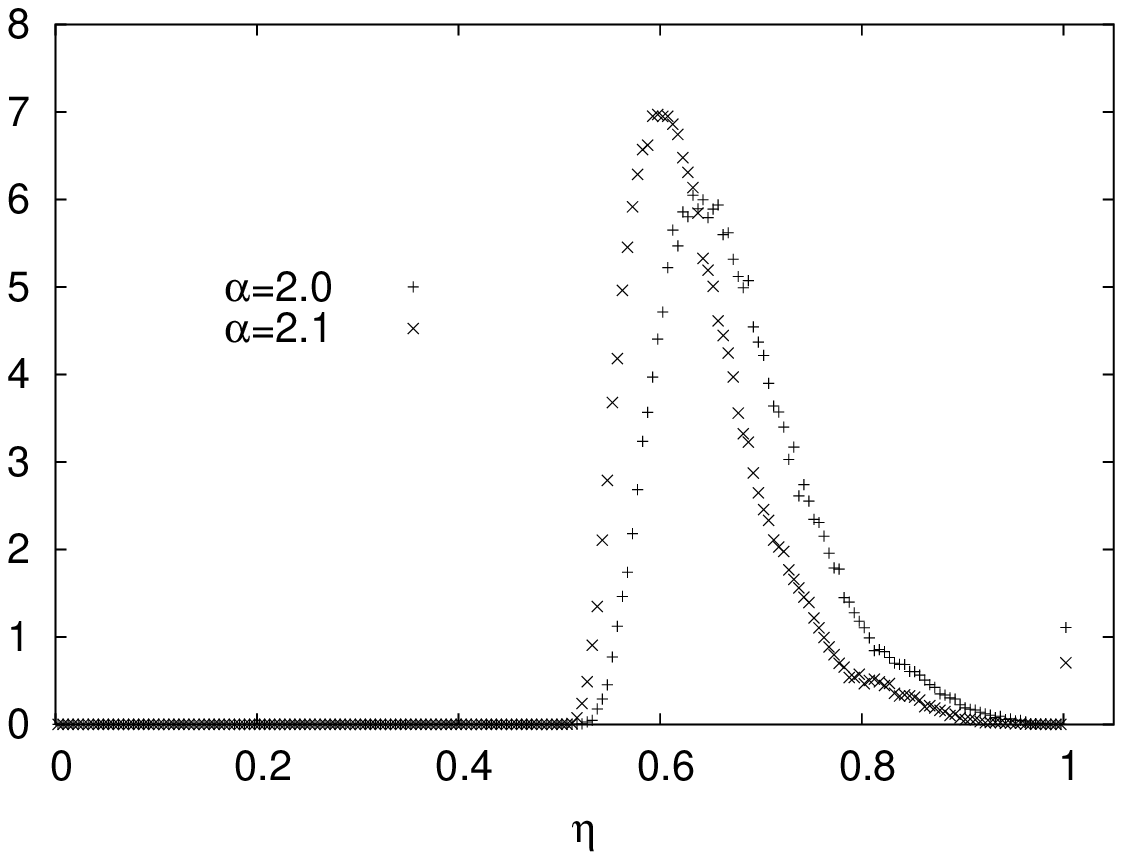}
\includegraphics[width=.49\textwidth]{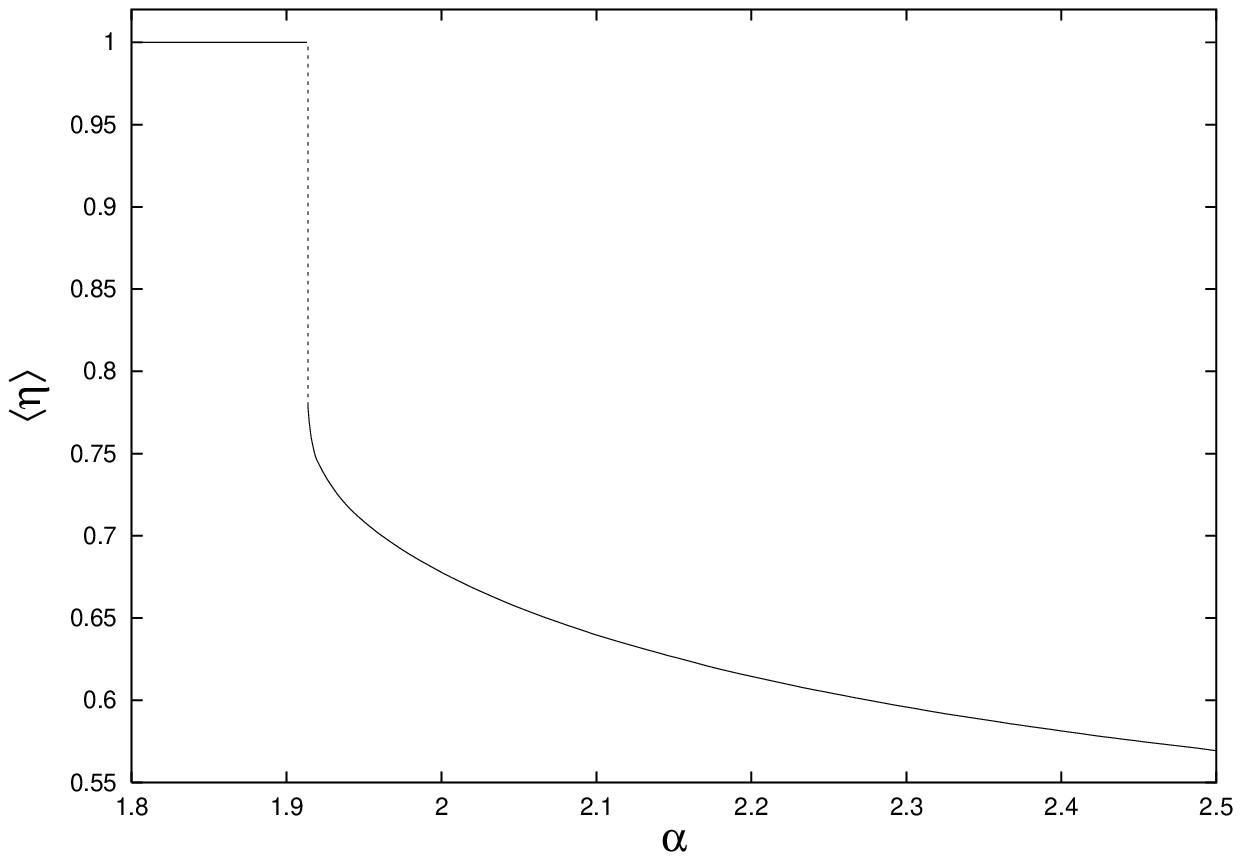}
\end{center}
\caption{Left: Probability distribution $\rho(\eta)$ for $\alpha=2.0$
and $\a=2.1$. Note the trivial contribution in 1.  Right: Average
value of $\eta$ versus $\alpha$. This value is exactly 1 for
$\alpha<\alpha_d=1.915$.}
\label{rhodieta}
\end{figure}

Self-consistency equations (\ref{selfc}) and (\ref{selfc2}) can be
written as a single self-consistency equation for the distribution
$\rho(\eta)$.  In the $\mu\rightarrow\infty$ limit it reads
\begin{equation} \label{rho}
\rho(\eta) = \sum_{k=0}^{\infty}f_{3\alpha}(k)
\sum_{k'=0}^{\infty}f_{3\alpha}(k') \int \prod_{i=1}^k d\rho(\eta_i)
\prod_{j=1}^{k'} d\rho(\eta'_j) \;\delta\left[\eta-1+\frac{1}{2}
\left(1-\frac{\prod_{i=1}^k\eta_i}{A_k}\right)
\left(1-\frac{\prod_{j=1}^{k'}\eta'_j}{A_{k'}}\right) \right]\;,
\end{equation}
with the normalization coefficients $A_k =
2\prod_{i=1}^{k}\frac{1+\eta_i}{2} - \prod_{i=1}^{k}\eta_i$.
Eq.(\ref{rho}) can be solved by a population dynamics algorithm.
Starting from a population of $\eta$s randomly distributed in $[0,1]$
we then iterate the following steps:
\begin{itemize}
\item take $k$ elements and compute $\eta^k$ and $A_k$, where $k$ is a
Poissonian number;
\item take $k'$ elements and compute $\eta^{k'}$ and $A_{k'}$, where
$k'$ is a Poissonian number too;
\item compute a new $\eta$ as
$$
1-\frac{1}{2} \left(1-\frac{\eta^k}{A_k}\right)
\left(1-\frac{\eta^{k'}}{A_{k'}}\right)\;,
$$
and insert it in the population eliminating another random $\eta$.
\end{itemize}
The asymptotic distribution $\rho(\eta)$ is plotted in figure
\ref{rhodieta} (left) for different values of $\alpha$.  For
$\alpha>\alpha_d\simeq 1.915$ the distribution has both a trivial
contribution in 1 and a non-trivial one in the $[\frac{1}{2};1]$
region, while for $\alpha<\alpha_d$ it collapses into a single delta
function in 1.

In figure \ref{rhodieta} (right) we plot the average value of $\eta$
versus $\alpha$, by which we immediately localize the dynamical phase
transition at $\alpha_d=1.915$. An identical curve has been calculated
analytically in the more tractable case of the $p$-spin model
\cite{pspin}.

\subsection{Complexity} 

In the $\mu\rightarrow\infty$ limit the complexity is given by
\cite{cavity}
\be
\Sigma = \lim_{\mu\rightarrow\infty}(-\mu\Phi) = \lim_{\mu\to\infty}
\left\{\overline{\log A_k} - 2\a \overline{\log
[1-\frac{1}{2}(1-\eta)(1-\frac{\eta^k}{A_k})]} \right\}
\ee
where the averages are taken with respect to the Poissonian distribution of
$k$ and with respect to $\rho(\eta)$.

\begin{figure}
\begin{center}
\includegraphics[height=.4\textwidth]{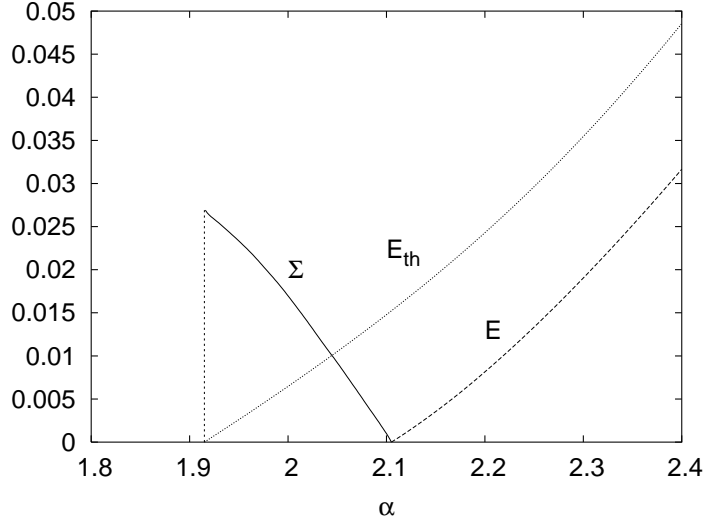}
\end{center}
\caption{Phase diagram of random hypergraph bicoloring.}
\label{dfase}
\end{figure}
  
The complexity curve is plotted in figure~\ref{dfase}: we identify the
critical point $\alpha_c=2.105$ that corresponds to the SAT/UNSAT
transition, as the point where the complexity vanishes.

\subsection{Energy density and 1RSB phase diagram}

In order to evaluate free energy $\Phi$ we must generalize the
computation to for finite values of $\mu$.
 
The self-consistency equation for general $\mu$ is 
\be
\rho(\eta) = \sum_{k=0}^{\infty}f_{3\alpha}(k)
\sum_{k'=0}^{\infty}f_{3\alpha}(k') \int \prod_{i=1}^k d\rho(\eta_i)
\prod_{j=1}^{k'} d\rho(\eta'_j)\; \delta\left[\eta-1+\frac{1}{2}
\left(1-\frac{a_k}{A_k}\right) \left(1-\frac{a_{k'}}{A_{k'}}\right)
\right]
\ee
where $a_k$ is the coefficient of the delta function in 0 of the
distribution $P^{(k)}(h)$ computed by the convolution of $k$ biases
$u$, and $A_k$ is its normalization factor.  To compute quickly the
$P^{(k)}(h)$ we can use a recursive relation:
\be
P^{(k)}(h)=\int dQ_k(u_k)\, dP^{(k-1)}(g)\, \delta(h-g-u_k)
e^{-\mu(|u_k|+|g|-|g+u_k|)} \;.
\ee
The free energy is given by $\Phi = \Phi_1-2\a\Phi_2$ with
\bea \label{delteegen}
\Phi_1 &=& -\frac{1}{\mu}\overline{\log(A_k)} \;, \nn \\
\Phi_2 &=& -\frac{1}{\mu}\overline{\log\left({1 -
\frac{1}{2}(1-\eta)(1-\frac{a_k}{A_k})}(1-e^{-2\mu}) \right)}\;.
\eea
For $\a>\a_c$, $\Phi$ has a maximum at a finite value of $\mu$: it
means that the ground state has positive energy.  Otherwise for
$\a<\a_c$ $\Phi$ is always negative, converging toward zero for
$\mu\rightarrow\infty$: it corresponds to a zero-energy ground state.

The energy density is calculated as
\be
E=\frac{\partial}{\partial\mu}(\mu\Phi)=-\frac{1}{A_k}\frac{\partial
A_k}{\partial\mu}+2\a \frac{(1-\eta)(1-\frac{a_k}{A_k})e^{-2\mu}}
{1-\frac{1}{2}(1-\eta)(1-\frac{a_k}{A_k})(1-e^{-2\mu})} \;.
\ee
As we did before, rather than computing the derivative of
the $A_k$, we can write a recursive equation for the probability
distribution $R^{(k)}(h)\equiv\frac{\partial}{\partial\mu}P^{(k)}(h)$:
\be
R^{(k)}(h) = \int dQ_k(u_k)\,dg[R^{(k-1)}(g) +
(|h|-|g|-|u_k|)P^{(k-1)}(g)] \delta(h-g-u_k)e^{-\mu(|g|+|u_k|-|h|)}\;.
\ee
Injecting this calculation in the population dynamics algorithm
provides directly the curve $E(\mu) = \frac{\partial}{\partial\mu}
(\mu\Phi)$.  The ground state energy is obtained as the point where
$E(\mu)$ and $\Phi(\mu)$ coincide.

\begin{figure} 
\begin{center}
\includegraphics[width=.5\textwidth]{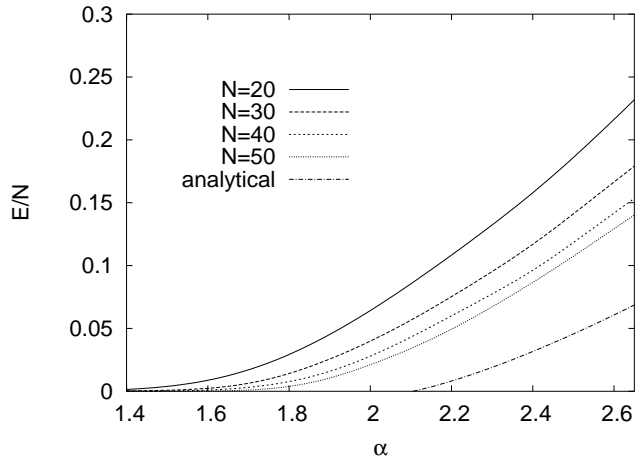}
\end{center}
\caption{Energy density of random hypergraph bicoloring: comparison
among finite-size numerical results and analytical 1RSB solution.}
\label{confronto}
\end{figure}

The ground state energy density is compared to the numerical results
in figure \ref{confronto}.  This curve must be considered a
$N\rightarrow\infty$ limit of the finite $N$ curves that we obtained
numerically.

\begin{figure}
\begin{center}
\includegraphics[width=.5\textwidth]{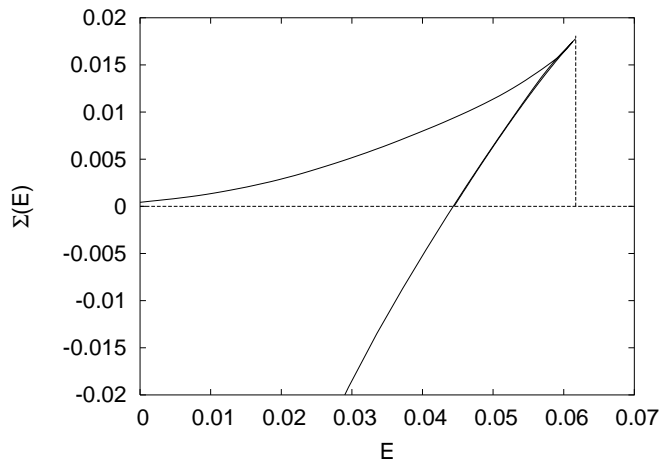}
\end{center}
\caption{Complexity versus energy at $\a=2.5$: note the non-physical
upper branch. Along the physical lower branch at the threshold energy
$E_{th}=0.062$ the complexity is maximal, while it becomes zero at the
ground state energy.}
\label{sigmadie}
\end{figure}

Another interesting curve that we can compute is the complexity versus
the energy, that we plot parametrically in $\mu$ using $E(\mu)$ and
$\Phi(\mu)$ (see Fig.~\ref{sigmadie}). The curve $\Sigma=\mu(E-\Phi)$
has two branches: the lower one is physical one and represents the
true complexity~\footnote{For the unphysical one there is not still a
precise interpretation \cite{cavity}, however it seems not to have any
physical meaning.}.

The last quantity we display in Fig.~\ref{dfase} is $E_{th}$ versus
$\a$, that is simply the maximum of $E(\mu)$.

Summarizing the 1RSB results we get the following scenario. 

There is a ``paramagnetic'' phase for $\a<\a_d=1.915$, where there are
no metastable states and we conjecture the existence of linear
algorithms for coloring the generic hypergraph. The cavity fields are
zero, so the spins are not forced to be black or white. In the
so-called HARD-SAT region $\a_d<\a<\a_c=2.105$ the generic hypergraph is
still colorable, but the presence of many states makes the coloring
procedure very difficult. In each ground state there is a core of
spins for which there is a particular pattern of coloring: because of
the existence of an exponentially larger number of metastable states,
it is very difficult for local search algorithm to color the core in
the right way. For $\a>\a_d$ the 1RSB approximation becomes less valid
when high energy states are considered \cite{fede+montan}. Most
likely, the curve $E_{th}$ would slightly change if a better
approximation would be used.

These 1RSB results are expected to be a very good approximation of the
exact analytical solution, as it happens in the majority of similar
combinatorial optimization problems. For the $p$-spin model
\cite{RIWEZE} an exact solution has been found that is identical to
the 1RSB one \cite{pspin,cocco}.
 

\section{Single sample analysis and the SP algorithm}

An innovative and useful reformulation of the cavity equations has
been proposed in ref. \cite{ksat}. The self-consistency equations are
used to study single random problem instances and allow to get a
microscopic information about the behavior of the single spins in the
stable and metastable states of given energy density. The method,
called Survey Propagation (SP), is general and provides the core
ingredient of a new efficient algorithm \cite{ksat,BMZ,MPZ} for
finding ground states within the glassy phase.  Here we will apply and
check SP for the bicoloring problem.  This problem is half-way between
the random K-SAT problem and the random K-XORSAT (or $p$-spin)
problem.  Since the SP algorithm does work for random K-SAT
\cite{ksat}, but it does not seem to work for random K-XORSAT, we
believe of primary importance to check its performances on the random
hypergraph bicoloring problem.

The iterative equations for the probability distributions of cavity
fields that we have used in the previous sections to find the phase
diagram were implementing at the same time a population dynamics
process and an averaging over the random realizations.  However, the
equations can be easily iterated over specific realizations, that is
avoiding the averaging step.  In such a formulation the order
parameter becomes the full list of the cavity fields over the entire
graph.  From the cavity fields one may determine the bias of each spin
in all metastable states of given energy density and this information
can be used for algorithmic purposes.  The underlying hypothesis for
the exactness of the single-sample formalism is the validity of the so
called clustering condition within states: cavity fields should be
uncorrelated within states and we expect this to be approximatively
true thanks to the fact that the most numerous loops in the graph have
a length that diverges as $\log N$.

\begin{figure}
\centering
\includegraphics[width=8cm]{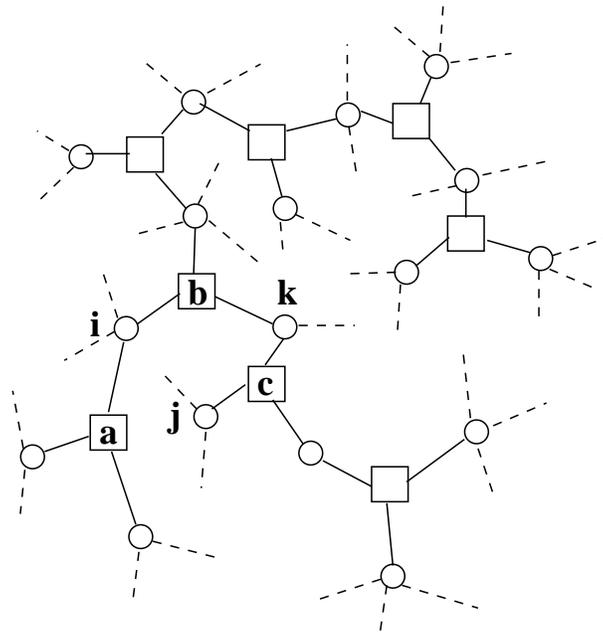}
\caption{Factor Graph representation of an energy minimization
problem.}
\label{fig:factor_graph}
\end{figure}

In order to set up an appropriate formalism for the single sample
analysis, we resort to the factor graph representation
\cite{factor_graph} of the bicoloring problem: variables are
represented by $N$ circular ``variable nodes'' labeled with letters
$i,j,k,...$ whereas links (which carry the interaction energy) are
represented by $M$ square ``function nodes'' labeled by
$a,b,c,...$(see Fig.~\ref{fig:factor_graph}).  Function nodes have
connectivity $3$, variable nodes have a Poisson connectivity of
average $3 \alpha$ and the overall graph is bipartite.  The energy
function can be trivially written as the sum over function nodes of
their energies.

\begin{figure}
\centering
\includegraphics[width=8cm]{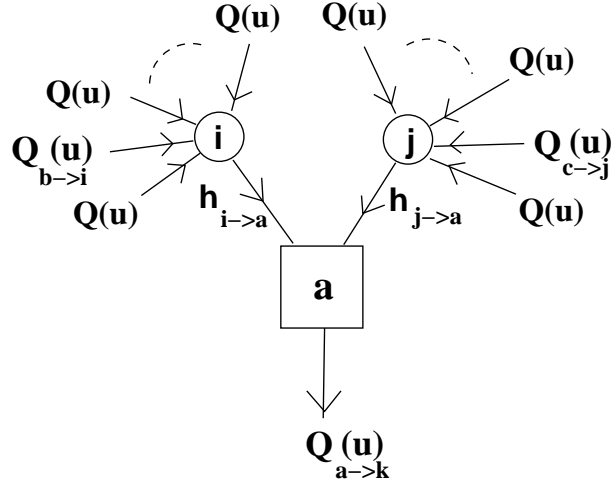}
\caption{Iterative equations as message-passing procedure.}
\label{fig:eq_Q}
\end{figure}

Following ref. \cite{ksat}, we call ``messages'' the $\hat u$ terms
which represent the contribution to the cavity fields coming from the
different connected branches of the graph.  In the message-passing
language (typical of error correcting codes algorithms \cite{Yedidia})
one may describe the SP equations as follows. In the replica symmetric
approximation, the messages arriving at a node are added up and then
sent to a function node. Next, the function node transforms all input
signals into a new message which is sent to the descendant variable
node.  At the 1RSB level, the messages along the links of the factor
graph are u-surveys of usual messages over the various possible states
of the system at a given value of the energy (which is fixed by the
reweighting parameter $\mu$).  While the method is not restricted to
zero temperature, at $T=0$ it assumes a particularly simple form
because messages can take only few values, $3$ in our case, and the
u-survey are given by the probabilities of these values. The u-surveys
are parametrized by $2$ real numbers and the SP can be implemented
easily.  Each edge $a \to j$ from a function node to a variable node
$j$ carries a u-survey $Q_{a \to j}(u)$. The algorithm finds these
u-surveys and all the cavity fields $P_{i \to a}(h)$. Very
schematically, the procedure works as follows.  All the u-surveys
$Q_{a \to i}(u)$ are initialized randomly. Next, function nodes are
selected sequentially at random and the u-surveys are updated
according the the equations:
\begin{equation}
P_{i \to a}(h )= C_{i \to a} \int du_1...du_{k} Q_{b_1 \to i}
(u_1)...Q_{b_{k} \to i}(u_{k}) \delta \left(h-\sum_{a=1}^{k} u_a
\right) \exp \left( \mu(\vert \sum_{a=1}^ku_a \vert - \sum_{a=1}^k
\vert u_a \vert) \right)
\label{P}
\end{equation}
\begin{equation}
Q_{a \to i}(u)=C_{a \to i} \int dg dh P_{j \to a}(g) P_{\ell \to a}(h)
\delta\left( u-\hat u_(g,h)\right)
\label{Q2}
\end{equation}
where the function $\hat{u}(g,h)$ is the one defined in
Eq.(\ref{def}).  In the above expressions, $C_{i \to a},C_{a \to i}$
are normalization constants and the labels $b_i$ identify the $k$
neighboring function nodes different from $a$ connected to site the
variable node $i$ (see Fig.~\ref{fig:eq_Q})

Parameterizing the u-surveys as 
\begin{equation}
Q_{a \to i}(u)=(1-\eta_{a \to i}^+-\eta_{a \to i}^-)\delta(u)+\eta_{a
\to i}^+\delta(u-1)+\eta_{a \to i}^-\delta(u+1)
\end{equation}
the above set of equations (\ref{P},\ref{Q2}) define a non-linear map
over the $\eta$s~\footnote{In the algorithmic formalism we need a more
general parametrization of surveys with respect to the one used in the
first sections. As we shall see, along the decimation process the
symmtries of surveys are lost.}.

\begin{figure}
\includegraphics[angle=0,width=0.65\columnwidth]{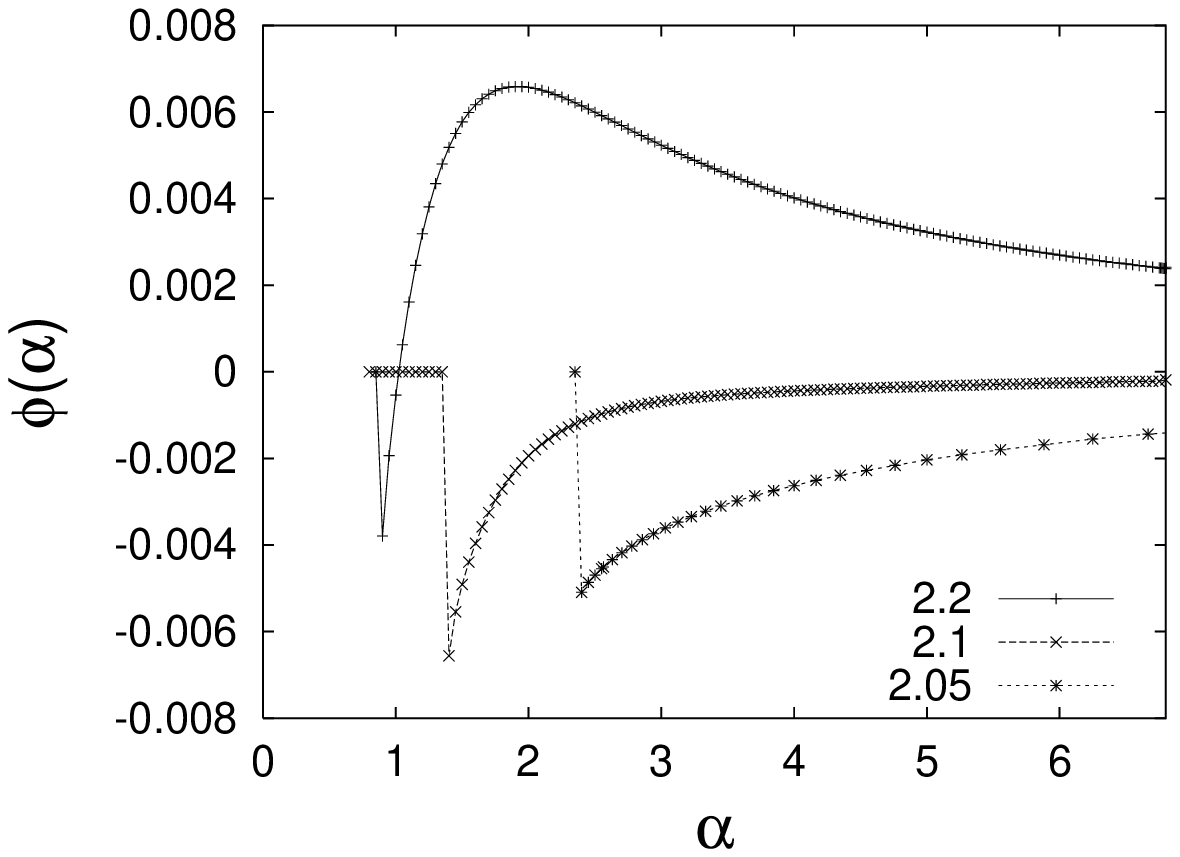}
\caption[0]{Free energy $\phi(\mu)$ for different samples of size
$N=10000$ and $\alpha=2.05,2.1,2.2$.}
\label{fig:free_energy}
\end{figure}  

The process is iterated until convergence is reached and finally the
stable set of u-surveys are used to compute the $N$ local field $\{
P_i(H_i)\})$ distributions and the free energy $\Phi(\mu)$. We have:
\begin{equation}
P_{i} (H)=C_i \int \prod_{a \in V(i)} du_a Q_{a \to i} (u_a) \delta
\left( H -\sum_{a \in V(i)} u_a \right) \exp\left(\mu ( \vert \sum_{a
\in V(i)} u_a \vert - \sum_{a \in V(i)} \vert u_a \vert \right)
\label{local_field}
\end{equation}
with $C_i$ being the normalization constant and $V(i)$ the set of
function nodes connected to variable $i$. The free-energy reads
\begin{equation}
\Phi(\mu)= \frac{1}{N}\left(\sum_{a=1}^M \Phi^f_{a}(\mu)-\sum_{i=1}^N
\Phi^v_i(\mu) (n_i-1)\right) \ ,
\label{free_onesamp1}
\end{equation}
where
\begin{eqnarray}
\Phi^f_{a}(\mu)&=&-\frac{1}{\mu} 
\log \left\{
\int \prod_{i \in V(a)}
\left[ \prod_{b\in V(i)-a} Q_{b\to i}(u_{b\to i}) d u_{b\to i} \right]
\right.
\nonumber
\\ 
\nonumber
&&\left.
\exp \left[ -\mu \min_{\{ \sigma_i,i \in V(a)\} } \left( E_a -\sum_{i
\in V(a)} \left[ \sum_{b\in V(i)-a}u_{b\to i} \right] \sigma_i +
\sum_{b\in V(i)-a} \vert u_{b\to i} \vert \right) \right] \right\}
\nonumber \\
\Phi^v_i(\mu)&=&-\frac{1}{\mu} \log \left \{ 
\int \prod_{a \in V(i)} du_a Q_{a \to i} (u_a)  \exp\left[\mu (\vert
\sum_{a \in V(i)} u_a \vert- \sum_{a \in V(i)}\vert u_a \vert )
\right] \right \}=-\frac{1}{\mu} \log (C_i)
\label{free_onesamp2}
\end{eqnarray}
In the above expressions, $V(a)$ identifies the set of variable nodes
connected to the function node $a$ and $E_a$ is its energy (i.e. the
link energy)..

\begin{figure}
\includegraphics[angle=0,width=0.65\columnwidth]{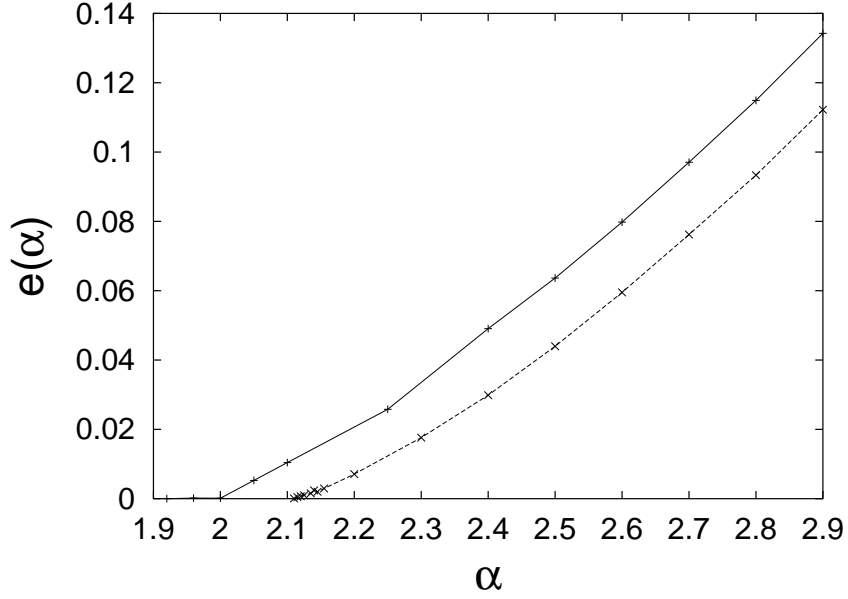}
\caption{Ground state energy and threshold energy for a single sample
of size $N=10000$ at different connectivities.}
\label{fig:energy}
\end{figure}  

The complexity $\Sigma(\mu)=\partial \Phi(\mu) / \partial(1/\mu)$ and
the energy density $\epsilon(\mu)=\partial(\mu\Phi(\mu))/\partial\mu$
of states can also be estimated over single instances.
Fig.~(\ref{fig:free_energy}) shows the free energy $\phi(\mu)$ of
single graphs with $N=10000$ vertexes as a function of $\mu$ for
different values of the average connectivity $\alpha$.
Fig.~(\ref{fig:energy}) shows the ground state energies and threshold
energies for single instances at different $\alpha$. Similar data can
be produced for the complexity. The agreement with the averaged
calculations of the previous sections is indeed remarkable already for
relatively small values of $N$ (as it should be expected from the
self-averaging property of the free-energy).

Once the information concerning the effective local fields acting on
the single spin variables becomes available a decimation procedure for
finding ground states can be easily implemented.  We have done one
such implementation for the $\mu \to \infty$ case, with the scope of
finding perfect colorings in the dynamical region just below
$\alpha_c$.  In this regime, the expression of the nonlinear map
simplifies considerably. From eqs. (\ref{P},\ref{Q2}) we find
\begin{eqnarray}
\eta^+_{a \to i}=\prod_{j \in V(a) \setminus i} \left[
\frac{\Pi^-_{j\to a}} {\Pi^0_{j\to a}+\Pi^-_{j\to a}+\Pi^0_{j\to a}}
\right] \ , \nonumber \\
\eta^-_{a \to i}=\prod_{j \in V(a) \setminus i} \left[
\frac{\Pi^+_{j\to a}} {\Pi^+_{j\to a}+\Pi^-_{j\to a}+\Pi^0_{j\to a}}
\right] \ , \label{eta1} 
\end{eqnarray}
where
\begin{eqnarray}
\Pi^+_{j \to a}&=& \prod_{b \in V(j) \setminus a}
\left(1-\eta^{-}_{b\to i}\right) - \prod_{b \in V(j) \setminus a}
\eta^{0}_{b\to i} \nonumber \\ 
\Pi^-_{j \to a}&=& \prod_{b \in V(j)
\setminus a } \left(1-\eta^{+}_{b\to i}\right)- \prod_{b \in V(j)
\setminus a } \eta^{0}_{b\to i} \nonumber \\ 
\Pi^0_{j \to a}&=&
\prod_{b \in V(j) \setminus a }\eta^{0}_{b\to i}
\label{pidef}
\end{eqnarray}
The value of $\eta ^0_{a\to j }$ can be calculated by
normalization. Other relevant quantities such as the biases of
variables and the complexity also acquire a simple form.  Upon
defining the bias $W_i^{\pm,0}$ of a variable as the probability of
picking up a cluster of ground states at random and find that variable
frozen in some preferential direction, that is $W_i^+ \equiv
\text{Prob}(H_i>0)$, $W_i^0\equiv\text{Prob}(H_i=0)$,
$W_i^-\equiv\text{Prob}(H_i<0)$, we have:
\begin{eqnarray}
W_i^{+} & = &
\frac{\hat\Pi_i^+}{\hat\Pi^+_{i}+\hat\Pi^-_{i}+\hat\Pi^0_{i}}
\nonumber \\
W_i^{-} & = & 
\frac{\hat\Pi_i^-}{\hat\Pi^+_{i}+\hat\Pi^-_{i}+\hat\Pi^0_{i}}
\nonumber \\
W_i^{0} & = & 1-W_i^{(+)}-W_i^{(-)} .
\label{wdef3}
\end{eqnarray}
with
\begin{eqnarray}
\hat \Pi^+_{i}&=& \prod_{ a\in V(i)}\left(1-\eta^{-}_{a\to i}\right) -
\prod_{ a} \eta^{0}_{a\to i} \nonumber \\
\hat \Pi^-_{i}&=& \prod_{ a\in V(i)}\left(1-\eta^{+}_{a\to i}\right)-
\prod_{ a} \eta^{0}_{a\to i} \nonumber \\
\hat \Pi^0_{i}&=& \prod_{a \in V(i)}\eta^{0}_{a\to i}
\label{hatpidef}
\end{eqnarray}
For the complexity we have:
\begin{equation}
\Sigma=\frac{1}{N} \left( \sum_{a=1}^M \Sigma_a -\sum_{i=1}^N
(n_i-1)\Sigma_i \right)
\label{sigma-def}
\end{equation}
where
\begin{equation}
\Sigma_a = \log \left[ \prod_{j \in V(a)} \left(\Pi^+_{j\to a} +
\Pi^-_{j\to a}+\Pi^0_{j\to a}\right)- \prod_{j \in V(a)}
\Pi^+_{j\to a} -\prod_{j \in V(a)}\Pi^-_{j\to a} \right]
\label{sigma_bond}
\end{equation}
\begin{equation}
\Sigma_i= \log\left[ \hat \Pi^+_{i}+ \hat \Pi^-_{i}+ \hat \Pi^0_{i}
\right]
\label{sigma_site}
\end{equation}

With the list of the biases on hand, the following simple decimation
procedure to find ground state configurations can been implemented:
{\tt
\begin{enumerate}
\item $\left\{ \eta\right\} \leftarrow$random
\item \label{2}SP
\begin{enumerate}
\item \label{a} Iterate eqs. (\ref{P},\ref{Q2}) until a fixed
  $\{\eta^*\}$ point is reached 
\end{enumerate}
\item Compute the biases 
$W_i^+=\text{Prob}(H_i>0)$, $W_i^0=\text{Prob}(H_i=0)$,
$W_i^-=\text{Prob}(H_i<0)$, following eq. (\ref{local_field}).
\item For $B_{i}=W_i^+-W_i^-$, Choose $i$ such that
$\left|B_{i}\right|$ is maximum.
\item IF $\left|B_{i}\right|<\epsilon$ for all $i$ then STOP
(paramagnetic state) and output the reduced sub-problem..
\item FIX $\sigma_{i}=1$ if $B_{i}>0$, $\sigma_{i}=-1$ otherwise.
\item GOTO \ref{2}
\end{enumerate}
}

One should notice that along the decimation procedure some of the
variable are fixed and therefore new types of links appear.  The
corresponding new function nodes will have an energy which is
inherited by the 3-body interaction by fixing one of the
variables. Once decimation has started, the bicoloring problems
becomes a mixture of graph and hypergraph bicoloring.

The behavior of the algorithm on sufficiently large ($n>10^3$) random
bicoloring instances is the following:
\begin{itemize}
\item for low $\alpha$ ($\alpha < \alpha_d$), the variables turn out
to be all paramagnetic (zero bias).
\item in the dynamical region the biases are non-trivial and the
decimation procedure fixes many variables leading to sub-problems
which are paramagnetic and easily solved by a greedy heuristic.  Very
close to $\alpha_c$ the decimation procedure may fail in finding
solutions in the first run.  In this region the algorithm can be
improved in many ways, e.g.\ by a random restart or a backtrack or a
different decimation strategy.  In any case we can not exclude the
existence of a threshold close to $\alpha_c$ where the decimation
procedure stops converging.
\end{itemize}

For small $N$ the structural ``rare events'' of the random
hyper-graph, like links sharing more than one variable or other types
of short loops, require an appropriate (in principle simple)
modification of the SP iterations \cite{Yedidia}.  More in general,
the presence of loops of different length scales may introduce
correlations which may require further non-trivial generalization of
the whole SP procedure.


\section{Conclusions}

In this work we have given a very detailed description of the random
hypergraph bicoloring problem, both on the average-case and on single
samples.

After having defined the statistical model corresponding to this
problem, we have applied the cavity method to solve it: results in the
RS and the 1RSB approximations have been presented.

Increasing the connectivity $\a$ the model undergoes several phase
transitions, which can be summarized as follows:
\begin{itemize}
\item for $\a<\a_s$ the model is in a genuine paramagnetic phase, all
the magnetizations are identically null;
\item at $\a=\a_s$ a ``replicon'' instability takes place, which
manifests at finite temperature with the onset of spin-glass order
(full RSB);
\item for $\a_s<\a<\a_d$ the presence of a full RSB phase at finite
temperatures is reflected in the ground states by finite values for
the spin magnetizations;
\item at $\a=\a_d$ a clustering transition takes place among the
ground states.  They split in an exponentially large number of
clusters. Within each cluster a finite fraction of variables is
completely frozen (backbone);
\item for $\a_d<\a<\a_c$ the model has a non-zero complexity and an
exponentially large number of metastable states, which may block
local-search algorithms. Although the very strong correlations among
variables the ground state energy is still zero and the problem is
colorable on average;
\item at $\a=\a_c$ the COL/UNCOL phase transition takes place;
\item for $\a>\a_c$ the ground state energy is positive and the
problem can not be colored on average.
\end{itemize}

In the second part of this work we have applied the Survey Propagation
algorithm to problem instances taken from the HARD-COL region
($\a_d<\a<\a_c$), finding in polynomial time solutions to the problem.
So we have verified that the SP algorithm works properly also for this
model, which is harder than the 3-sat problem~\cite{ksat}.  Indeed
this model, at variance with K-SAT, has no local biases which could in
principle be exploited by a smart algorithm.  

Next steps in this line of research will be to consider random hard
combinatorial problems endowed with some non trivial local structure
of the underlying graph. This constitutes a conceptual challange that
will bring the algorithmic and anlytical tools developed for sparse
graphs closer to what is found in the real-world version of the same
class of models\cite{SATLIB}.

\end{document}